\documentclass[12pt]{article}

\usepackage{float}
\usepackage{amsfonts}

\usepackage{amssymb}
\usepackage{latexsym}
\usepackage{graphicx}
\usepackage[english]{babel}
\usepackage[font={small}]{caption}

\usepackage{amsfonts}
\usepackage{latexsym}
\usepackage{graphicx}
\usepackage[english]{babel}
\usepackage{tikz}

\begin{document}
\input epsf

\def\p{\partial}
\def\h{{1\over 2}}
\def\be{\begin{equation}}
\def\bea{\begin{eqnarray}}
\def\ee{\end{equation}}
\def\eea{\end{eqnarray}}
\def\d{\partial}
\def\la{\lambda}
\def\eps{\epsilon}
\def\bb{\bigskip}
\def\mm{\medskip}
\newcommand{\dm}{\begin{displaymath}}
\newcommand{\edm}{\end{displaymath}}
\renewcommand{\b}{\tilde{B}}
\newcommand{\gm}{\Gamma}
\newcommand{\ac}[2]{\ensuremath{\{ #1, #2 \}}}
\renewcommand{\ell}{l}
\newcommand{\z}{\ell}
\newcommand{\newsection}[1]{\section{#1} \setcounter{equation}{0}}
\def\bb{$\bullet$}
\def\Qbar{{\bar Q}_1}
\def\QPbar{{\bar Q}_p}

\def\q{\quad}

\def\bn{B_\circ}

\let\a=\alpha \let\b=\beta \let\g=\gamma \let\d=\delta \let\e=\epsilon
\let\c=\chi \let\th=\theta  \let\k=\kappa
\let\l=\lambda \let\m=\mu \let\n=\nu \let\x=\xi \let\r=\rho
\let\s=\sigma \let\t=\tau
\let\vp=\varphi \let\vep=\varepsilon
\let\w=\omega      \let\G=\Gamma \let\D=\Delta \let\Th=\Theta
                     \let\P=\Pi \let\S=\Sigma

\def\h{{1\over 2}}
\def\t{\tilde}
\def\r{\rightarrow}
\def\nn{\nonumber\\}
\let\bm=\bibitem
\def\Kt{{\tilde K}}
\def\b{\bigskip}

\let\p=\partial

\begin{flushright}
\end{flushright}
\vspace{20mm}
\begin{center}
{\LARGE Can the universe be described by a wavefunction?  \footnote{Essay awarded an honorable mention in the Gravity Research Foundation 2018 Awards for Essays on Gravitation.}}
\\
\vspace{18mm}
 Samir D. Mathur

\vskip .1 in

 Department of Physics\\The Ohio State University\\ Columbus,
OH 43210, USA\\mathur.16@osu.edu\\
\vspace{4mm}
 March 31, 2018
\end{center}
\vspace{10mm}
\thispagestyle{empty}
\begin{abstract}

Suppose we assume that in gently curved spacetime (a) causality is not violated to leading order  (b) the Birkoff theorem holds to leading order and (c) CPT invariance holds. Then we argue that the `mostly empty' universe we observe around us cannot be described by an exact wavefunction $\Psi$. Rather, the weakly coupled particles we see are approximate quasiparticles arising as excitations of a `fuzz'. The `fuzz'  {\it does} have an exact wavefunction $\Psi_{fuzz}$, but this exact wavefunction does not directly describe local particles. The argument proceeds by relating the cosmological setting  to the black hole information paradox, and then using the small corrections theorem to show the impossibility  of an exact wavefunction describing the visible universe.

\end{abstract}
\vskip 1.0 true in

\newpage
\setcounter{page}{1}

\begin{figure}[h]
\begin{center}
 \includegraphics[scale=1.] {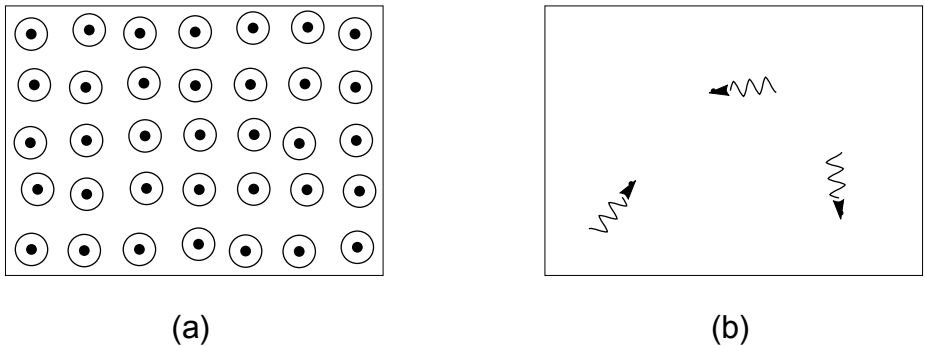}
\end{center}
\caption{ (a) The complicated exact description of a crystal  (b)  An approximate effective description of the dynamics in terms of phonons. } 
\label{f0}
\end{figure}

Consider a crystal. The lattice vibrations may be quantized to yield phonons. If we focus on the low energy excitations of the crystal, we  see a set of weakly interact phonons, so we might picture the physics as  in fig.\ref{f0}(b). But this dynamics is an approximation; the exact  wavefunction of the crystal is a  complicated one involving quarks, gluons and leptons (fig.\ref{f0}(a)). 

We will argue that the information paradox implies a similar situation for our universe. The particles we see around us are like the phonons of fig.\ref{f0}(b), while the exact wavefunction of the universe describes a very different set of degrees of freedom. 

\b

First we recall the rigorous formulation of the information paradox \cite{cern}. Hawking \cite{hawking} showed that entangled pairs are created at the horizon, which leads to a monotonically growing entanglement between the emitted radiation and the remaining hole. This leads to a problem near the endpoint of evaporation: we  lose unitarity if the hole disappears, or we end up with a planck sized remnant having infinitely many possible internal states. In string theory we must maintain unitarity; further AdS/CFT duality rules out remnants as the CFT has finitely many states for bounded energy. 

Many string theorists harbored the hope that the problem would be resolved by subtle corrections of order $\epsilon\ll 1$ to the state of each emitted pair; since there are many such pairs, the overall entanglement might vanish by the endpoint of evaporation. But in \cite{cern} it was shown, using the strong subadditivity of quantum entanglement entropy, that the entanglement $S_{ent}(N)$ after $N$ emissions keeps rising
\be
S_{ent}(N+1)>S_{ent}(N)+\ln 2 -2\epsilon
\label{one}
\ee
Thus we cannot have the standard semiclassical picture of the black hole as a leading order approximation; there must be an {\it order unity} change at the horizon.
In string theory we indeed find the microstates of the hole are horizon sized {\it fuzzballs} with no horizon; such states radiate from their surface like normal warm body, and there is no information paradox \cite{fuzzballs}.

Now consider the collapse of a homogeneous dust ball of mass $M$ and radius $R(\tau)$. We state our assumptions and their implications:

\b

\begin{figure}[h]
\begin{center}
 \includegraphics[scale=1] {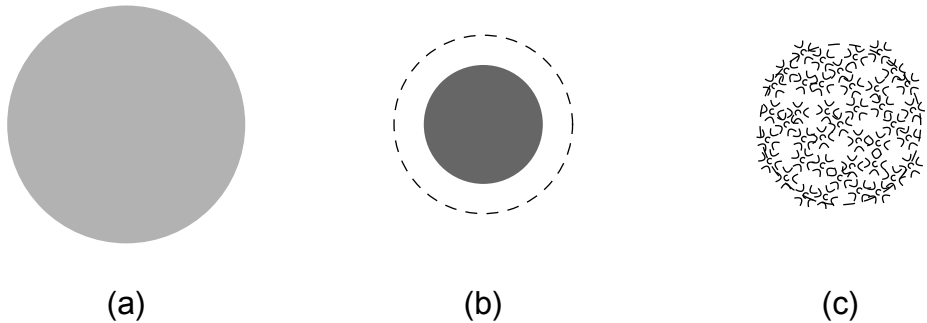}
\end{center}
\caption{ (a) A collapsing dust ball (b) The ball should never reach the semiclassically expected configuration (b), else we cannot solve the information puzzle (c) In string theory fuzzballs form instead when the ball reaches horizon radius. } 
\label{f1}
\end{figure}

(A) We assume that {\it causality holds to leading order in gently curved spacetime}.  Then we can never have the situation in fig.\ref{f1}(b), where the ball has collapsed inside $R_s=2GM$ to leave a smooth horizon. Indeed, were such a horizon to form, then we cannot escape the information paradox: information cannot come out of the hole at leading order, and the entanglement of pairs also keeps growing at leading order.  By (\ref{one}),  the corrections to the field dynamics arising from small violations of causality cannot overwhelm the relentless growth of entanglement.  We are forced to conclude that  gravitational collapse must halt before $R(\tau)$ reaches $R_s$. In string theory this indeed happens \cite{tunnel, causality}: a very large phase space corresponding to $Exp[S_{bek}[M]] $ fuzzballs opens up as $R\r R_s$, and the wavefunction of the ball spreads over this space instead of continuing on its semiclassical trajectory to $R<R_s$ (fig.\ref{f1}(c)). 

\b
 
\begin{figure}[H]
\begin{center}
 \includegraphics[scale=.8] {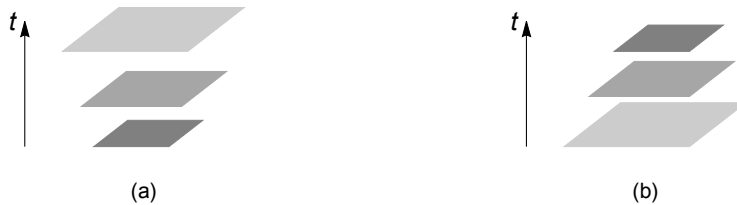}
\end{center}
\caption{(a) An expanding universe can be mapped by CPT to  (b) a collapsing universe. } 
\label{f2}
\end{figure}

(B) Now consider a flat dust cosmology
\be
ds^2=-dt^2 + a^2(t) [ dr^2+r^2 d\Omega_2^2]
\label{two}
\ee
where $a(t)=a_0 t^{2\over 3}$. We assume that {\it CPT invariance holds}. Then we can map this expanding universe (fig.\ref{f2}(a)) to a collapsing one (fig.\ref{f2}(b)).

\b

\begin{figure}[H]
\begin{center}
 \includegraphics[scale=1.1] {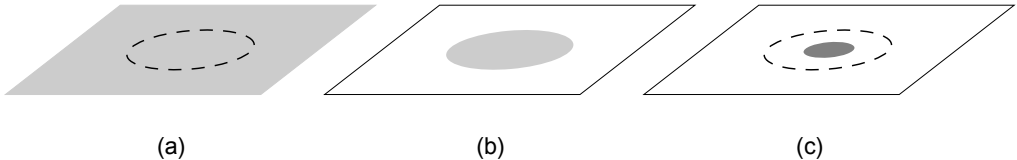}
\end{center}
\caption{ (a) A homogeneous collapsing flat cosmology (b) By the Birkoff theorem,  the dynamics of the marked ball cannot change if we remove the matter outside (c) Semiclassical dynamics suggests that this ball will pass through its horizon radius, but in actual fact the ball must tunnel to fuzzballs before this.} 
\label{f3}
\end{figure}

(C) Assume that {\it the Birkoff theorem holds to  leading order even in the full quantum theory.} Consider a ball of ball of proper radius $R(t)$ in the collapsing cosmology.  Then we can  replace the dust in the exterior region $R>R(t)$ by flat spacetime; by the Birkoff theorem this should not affect the dynamics of the ball in the region $R<R(t)$ (fig.\ref{f3}(a,b)). Note that in a dust cosmology there is no pressure across the surface $R(t)$; this is important to allow the separation  of the outer and inner regions in this manner. 

\b

\b

(D) In the semiclassical picture of collapse, this dust ball will pass through its horizon at some time $t$ to a radius satisfying $R<2GM$. If we choose our initial ball large enough, then this happens when the dust density is still very low compared to planck (fig.\ref{f3}(c)). 

\b

(E) But in (A) we argued that we cannot reach a situation where a horizon forms; instead, the dust ball must tunnel into a very quantum state of fuzzballs before it reaches its horizon radius $R_s$. {\it Thus the traditional picture of the cosmology where (\ref{two}) is a good semiclassical approximation describing particles of dust cannot be correct.}

\b

We have reached a startling conclusion based on very simple arguments, so let us pinpoint the crux of the logic. Steps (B), (D)   are standard; in fact it is often said that we live inside a white hole, which is just the time reverse of a black hole. Quite separately, we have always had (A), the information paradox, whose resolution we have not known until recent progress with string theory. The link between the two is (C), where the Birkoff theorem is used to discard the outside of a collapsing region, and thus convert a collapsing cosmology to the  collapsing dust ball of the black hole problem. 

One might argue that the Birkoff theorem is classical, so there can be small quantum corrections to this theorem. Such corrections could allow the exterior of our dust ball to have some small quantum effects on the interior, and change cosmological evolution in the  region $R<R(\tau)$  slightly away from what we had in the black hole problem. {\it But the small corrections theorem (\ref{one}) tells us that such small corrections cannot get us out of the information paradox.} Thus we can still use the Birkoff theorem to argue that just as we should not form a black hole horizon,  we should not get a {\it cosmological} horizon either. 

To summarize the argument, if a collapsing dust ball has to tunnel to fuzzballs when it reaches horizon radius, it should do that regardless of whether it is sitting in asymptotically flat spacetime or as part of a homogeneous isotropic cosmology. 

\begin{figure}
\begin{center}
 \includegraphics[scale=.7] {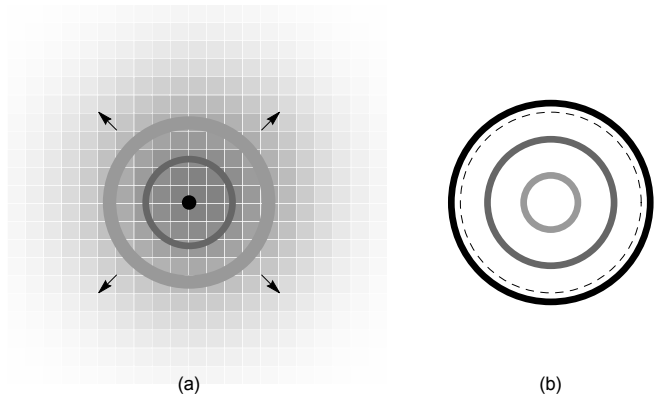}
\end{center}
\caption{ (a) A normal matter shell of horizon radius is the state represented by the dot in the center; this evolves to a linear combination of fuzzballs states, with the state moving out in superspace towards more complicated fuzzballs. (b) This evolution can be mapped, {\it approximately}, to the semiclassical infall of the shell  into a region $R<R_s$.} 
\label{f4}
\end{figure}

We cannot see past our Cosmlolgical horizon today. But astronomical evidence indicates that the horizon was smaller in the past, and the universe encompassed a dust ball much larger than horizon size.  To reconcile this with our claim above, we are forced   to the picture of `fuzzball complementarity' \cite{fuzzcomp,causality}, which gives rise to  an {\it approximate} emergence of semiclassical physics in interior of a fuzzball. In this picture, a collapsing dust ball tunnels into a linear combination of fuzzballs as its radius reaches the Schwarzschild radius, so we never really get a vacuum region inside a horizon. But evolution continues in {\it superspace} --  the very large space formed by the $Exp[S_{bek}]$ fuzzball states. This superspace is like our crystal, and wwavefunctions on superspace are like phonons. Evolution of these wavefunctions in superspace can me mapped, {\it in an approximate way} to evolution of the semiclassical dynamics of infalling quanta in a semiclassical picture of the black hole interior (fig.\ref{f4}). 

\b

The crucial fact here is the word {\it approximate}: semiclassical infall  is obtained for quanta with energy $E\gg T$ where $T$ is the temperature of the black hole. The information and entanglement of the hole is carried by Hawking quanta with energy $E\sim T$, and it is crucial that the approximation {\it fails} for such modes; else the semiclassical computation of Hawking would again be recovered, and (\ref{one}) would preclude any resolution of the information paradox. For black hole $T\sim 1/R_s$, and the errors in the complementary description are 
\be
\sim ~\left ( {E\over T}\right ) ^{1\over D-2}
\ee
where $D$ is the spacetime dimension. Thus for our 4-dimensional cosmology, we expect corrections of order
\be
(E R_H)^\h
\ee
where $E$ is the local energy of the quantum being studied, and $R_H$ is the radius of the cosmological horizon.

We thus see that such corrections will be too small to be measured today. But what is important is the question of principle: the traditional picture of a dust cosmology does not correspond to an exact wavefunction; the actual exact wavefunction is very different. To conclude, we note the structure of the exact wavefunction $\Psi$ and its approximations.

Let the collapsing cosmology start with a large ball of mass $M$ and radius $R$, with $R>R_s=2GM$.  The ball collapses semiclassically to size $R\approx R_s$, and then tunnels into fuzzballs. The evolution in superspace -- describing the collective excitations of the fuzzball -- maps to an {\it approximate} description of semiclassically moving dust particles that continue their collapse to $R<R_s$. But this semiclassical approximation can be applied only to a small patch at a time -- a patch smaller that horizon size. 

\begin{figure}
\begin{center}
 \includegraphics[scale=.6] {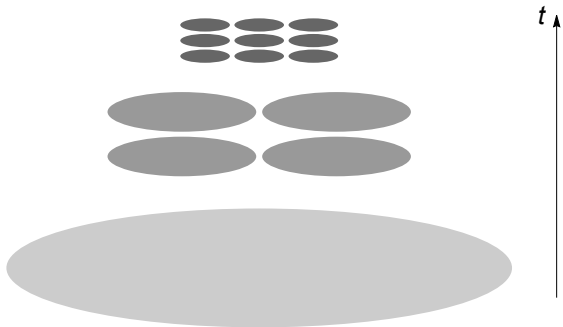}
\end{center}
\caption{A collapsing dust ball stays at its horizon radius, but approximate effective semiclassical infall can be obtained (through fuzzball complementarity) for  small patches. When these smaller patches reach their horizon radius,  we can get an  approximate complementary description of an even smaller patch etc.} 
\label{f5}
\end{figure}

At some point this smaller patch tries to enter its own horizon $R'_s$, at which point the particles obtained in the approximate semiclassical description stop behaving like free particles and make fuzzballs  of radius $R'_s$ instead.   We must now focus on a smaller `subpatch' to get a semiclassical evolution, and so on  (fig.\ref{f5}).  Our expanding cosmology is obtained as a time reverse of this hierarchical set of effective approximate descriptions.

We have been forced to the conclusion that `empty space' should have a highly entropic structure, computable in principle using the fuzzball construction. Low energy physics emerges as an effective approximate dynamics of collective modes.  There is no cosmological singularity:   we are forced to ever smaller patches of effective semiclassical physics as we approach the singularity, while the exact wavefunction describes the evolution of a huge fuzzball. 

\b

\section*{Acknowledgements}

I am grateful to Patrick Dasgupta, Oleg Lunin and Anupam Majumdar for helpful discussions. This work is supported in part by a grant from the FQXi foundation.

\end{document}